\documentclass[epjCONF,columns]{svjour} 
\usepackage{graphics}
\usepackage[varg]{txfonts} 
\usepackage[latin1]{inputenc}
\usepackage{ifthen} 

\usepackage{subfigure} 
\usepackage{rotating}

\usepackage{lineno}  
\usepackage{graphicx}  

\usepackage{xspace}  
\usepackage{color} %
\usepackage{colortbl} %
\usepackage{ctable}
\usepackage{upgreek}
\newboolean{uprightparticles}
\setboolean{uprightparticles}{false} 

\def\lhcb {LHCb\xspace}

\ifthenelse{\boolean{uprightparticles}}%
{
 
 \def\Pgamma      {\ensuremath{\upgamma}\xspace}

 \def\Pmu         {\ensuremath{\upmu}\xspace}

 \def\Ppsi        {\ensuremath{\uppsi}\xspace}

 \def\PDelta      {\ensuremath{\Delta}\xspace}                 
 \def\PXi      {\ensuremath{\Xi}\xspace}                 
 \def\PLambda      {\ensuremath{\Lambda}\xspace}                 
 \def\PSigma      {\ensuremath{\Sigma}\xspace}                 
 \def\POmega      {\ensuremath{\Omega}\xspace}                 
 \def\PUpsilon      {\ensuremath{\Upsilon}\xspace}                 
 
 \def\PB      {\ensuremath{\mathrm{B}}\xspace}                 
                  
 \def\PD      {\ensuremath{\mathrm{D}}\xspace}

 \def\PJ      {\ensuremath{\mathrm{J}}\xspace}                 
 \def\PK      {\ensuremath{\mathrm{K}}\xspace}

 \def\PW      {\ensuremath{\mathrm{W}}\xspace}

 \def\Pb      {\ensuremath{\mathrm{b}}\xspace}

 \def\Pi      {\ensuremath{\mathrm{i}}\xspace}

}
{
 
 \def\Pgamma      {\ensuremath{\gamma}\xspace}

 \def\Pmu         {\ensuremath{\mu}\xspace}

 \def\Ppsi        {\ensuremath{\psi}\xspace}                 
                  
 \mathchardef\PDelta="7101
 \mathchardef\PXi="7104
 \mathchardef\PLambda="7103
 \mathchardef\PSigma="7106
 \mathchardef\POmega="710A
 \mathchardef\PUpsilon="7107
                  
 \def\PB      {\ensuremath{B}\xspace}                 
                  
 \def\PD      {\ensuremath{D}\xspace}

 \def\PJ      {\ensuremath{J}\xspace}                 
 \def\PK      {\ensuremath{K}\xspace}

 \def\PW      {\ensuremath{W}\xspace}

 \def\Pb      {\ensuremath{b}\xspace}

 \def\Pi      {\ensuremath{i}\xspace}

}

\def\mup        {\ensuremath{\Pmu^+}\xspace}
\def\mun        {\ensuremath{\Pmu^-}\xspace} 
\def\mumu       {\ensuremath{\Pmu^+\Pmu^-}\xspace}

\def\g      {\ensuremath{\Pgamma}\xspace}

\def\W      {\ensuremath{\PW}\xspace}

\def\b     {\ensuremath{\Pb}\xspace}

\def\kaon  {\ensuremath{\PK}\xspace}
  \def\Kbar  {\kern 0.2em\overline{\kern -0.2em \PK}{}\xspace}

\def\Kz    {\ensuremath{\kaon^0}\xspace}
\def\Kzb   {\ensuremath{\Kbar^0}\xspace}
\def\KzKzb {\ensuremath{\Kz \kern -0.16em \Kzb}\xspace}
\def\Kp    {\ensuremath{\kaon^+}\xspace}
\def\Km    {\ensuremath{\kaon^-}\xspace}

\def\KpKm  {\ensuremath{\Kp \kern -0.16em \Km}\xspace}

\def\Kstarz  {\ensuremath{\kaon^{*0}}\xspace}

  \def\Dbar    {\kern 0.2em\overline{\kern -0.2em \PD}{}\xspace}
\def\D       {\ensuremath{\PD}\xspace}

\def\Dz      {\ensuremath{\D^0}\xspace}
\def\Dzb     {\ensuremath{\Dbar^0}\xspace}
\def\DzDzb   {\ensuremath{\Dz {\kern -0.16em \Dzb}}\xspace}
\def\Dp      {\ensuremath{\D^+}\xspace}
\def\Dm      {\ensuremath{\D^-}\xspace}

\def\DpDm    {\ensuremath{\Dp {\kern -0.16em \Dm}}\xspace}

\def\B       {\ensuremath{\PB}\xspace}
  \def\Bbar    {\kern 0.18em\overline{\kern -0.18em \PB}{}\xspace}

\def\Bd      {\ensuremath{\B^0}\xspace}
\def\Bs      {\ensuremath{\B^0_s}\xspace}

\def\jpsi     {\ensuremath{{\PJ\mskip -3mu/\mskip -2mu\Ppsi\mskip 2mu}}\xspace}

  \def\Y#1S{\ensuremath{\PUpsilon{(#1S)}}\xspace}

\newcommand{\tev}{\ensuremath{\mathrm{\,Te\kern -0.1em V}}\xspace}
\newcommand{\gev}{\ensuremath{\mathrm{\,Ge\kern -0.1em V}}\xspace}
\newcommand{\mev}{\ensuremath{\mathrm{\,Me\kern -0.1em V}}\xspace}
\newcommand{\kev}{\ensuremath{\mathrm{\,ke\kern -0.1em V}}\xspace}
\newcommand{\ev}{\ensuremath{\mathrm{\,e\kern -0.1em V}}\xspace}
\newcommand{\gevc}{\ensuremath{{\mathrm{\,Ge\kern -0.1em V\!/}c}}\xspace}
\newcommand{\mevc}{\ensuremath{{\mathrm{\,Me\kern -0.1em V\!/}c}}\xspace}
\newcommand{\gevcc}{\ensuremath{{\mathrm{\,Ge\kern -0.1em V\!/}c^2}}\xspace}
\newcommand{\gevgevcccc}{\ensuremath{{\mathrm{\,Ge\kern -0.1em V^2\!/}c^4}}\xspace}
\newcommand{\mevcc}{\ensuremath{{\mathrm{\,Me\kern -0.1em V\!/}c^2}}\xspace}

\def\invpb {\ensuremath{\mbox{\,pb}^{-1}}\xspace}

\def\invfb   {\ensuremath{\mbox{\,fb}^{-1}}\xspace}

\def\BR         {{\ensuremath{\cal B}\xspace}}

\newcommand{\decay}[2]{\ensuremath{#1\!\to #2}\xspace}         

\def\to                 {\ensuremath{\rightarrow}\xspace}

\def\CP                {\ensuremath{C\!P}\xspace}

\def\gsim{{~\raise.15em\hbox{$>$}\kern-.85em
          \lower.35em\hbox{$\sim$}~}\xspace}
\def\lsim{{~\raise.15em\hbox{$<$}\kern-.85em
          \lower.35em\hbox{$\sim$}~}\xspace}

\def\BdToKstmm    {\decay{\Bd}{\Kstarz\mup\mun}\xspace}

\def\BdToJPsiKst  {\decay{\Bd}{\jpsi\Kstarz}\xspace}

\def\BsPhiGam     {\decay{\Bd}{\phi \g}\xspace}
\def\BdKstGam     {\decay{\Bd}{\Kstarz \g}\xspace}

\def\AT#1     {\ensuremath{A_T^{#1}}\xspace}           

\def\Bsmm     {\decay{\Bs}{\mup\mun}\xspace}
\def\Bdmm     {\decay{\Bd}{\mup\mun}\xspace}

\def\C#1      {\ensuremath{\mathcal{C}_{#1}}}                       
\def\Cp#1     {\ensuremath{\mathcal{C}_{#1}^{'}}}                    
\def\Ceff#1   {\ensuremath{\mathcal{C}_{#1}^{\mathrm{(eff)}}}}        
\def\Cpeff#1  {\ensuremath{\mathcal{C}_{#1}^{'\mathrm{(eff)}}}}       
\def\Ope#1    {\ensuremath{\mathcal{O}_{#1}}}                       
\def\Opep#1   {\ensuremath{\mathcal{O}_{#1}^{'}}}                    

\newcommand{\CLs}{\ensuremath{\textrm{CL}_{\textrm{s}}}\xspace}

\newcommand{\Bsmumu}{\ensuremath{\Bs\to\mu^+\mu^-}\xspace}
\newcommand{\Bdmumu}{\ensuremath{\Bd\to\mu^+\mu^-}\xspace}
\newcommand{\Bhh}{\ensuremath{B^0_{(s)}\to h^+h^{'-}}\xspace}

\newcommand{\BuJpsiK}{\ensuremath{B^+\to J/\psi K^+}\xspace}

\newcommand{\BdJpsiKst}{\ensuremath{B^0_d\to J/\psi K^{*0}}\xspace}

\newcommand{\Bqmumu}{\ensuremath{\ensuremath{B^0_{(s)}}\to\mu^+\mu^-}\xspace}

\newcommand{\Lambdappi}{\ensuremath{\Lambda\to p\pi^-}\xspace}

\newcommand{\BRof}[1]{\ensuremath{{\cal B}(#1)}\xspace}

\newcommand{\figref}[1]{Fig.~\ref{#1}}

\newcommand{\secref}[1]{Sect.~\ref{#1}}

\def\BdToKstmm    {\decay{\Bd}{\Kstarz\mup\mun}\xspace}
\def\BdKstG {\decay{\Bd}{\Kstarz \g}\xspace}
\def\BsPhiG {\decay{\Bs}{\phi \g}\xspace}
\newcommand{\BRBdKstGam}{\ensuremath{\BRof\BdKstG\xspace}}
\newcommand{\BRBsPhiGam}{\ensuremath{\BRof\BsPhiG\xspace}}
\session-title{Hadron Collider Physics Symposium 2011, November 14-18, Paris, France}
\begin{document}
\title{Rare Decays in LHCb}
\author{Diego Mart{\'i}nez Santos\fnmsep\thanks{\email{diego.martinez.santos@cern.ch}}, on behalf of the LHCb Collaboration}
\institute{European Organisation for Nuclear Research (CERN), Geneva, Switzerland.}
\abstract{
The rare $B$ decays \Bqmumu,  \BdToKstmm and \BsPhiG  are studied
using up to $\sim 0.41$ \invfb of $pp$ collisions at $\sqrt{s}$ = 7~TeV
collected by the \lhcb experiment in 2010 and 2011.
A search for the decays \Bqmumu is performed with $0.41$ \invfb . The absence of significant signal leads to 
\BRof \Bsmumu $<1.4 \times 10^{-8}$ and \BRof \Bdmumu $<3.2 \times 10^{-9}$ at 95\,\% 
confidence level.  The forward-backward asymmetry, fraction of longitudinal polarization and differential branching 
fraction of \BdToKstmm, as a function of dimuon invariant mass, are measured in $0.31$ \invfb. 
 The ratio of branching ratios of the radiative \B decays \BdKstG and \BsPhiG has been measured using $0.34$ \invfb. The obtained value for the ratio 
is $ 1.52 \pm 0.14 \mathrm{(stat)} \pm 0.10 \mathrm{(syst)} \pm 0.12 (f_s/f_d)$. Using the HFAG value for \BRof\BdKstG, \BRof\BsPhiG has been found to be $(2.8 \pm 0.5 )\times 10^{-5}$.
} 
\maketitle
\section{Introduction}
\label{intro}
The \lhcb experiment ~\cite{lhcb} has provided preliminary results in the measurement of the forward-backward asymmetry, fraction of longitudinal polarization and differential branching 
fraction of \BdToKstmm ~\cite{kstmm} and the measurement of the \BRof\BsPhiG ~\cite{phiG}. 
\lhcb has also provided  upper limits in \BRof\Bsmumu and \BRof\Bdmumu ~\cite{bsmm}. 
\secref{sec:kstmm} sumarizes the analysis and results obtained by \lhcb in the study of \BdToKstmm. \secref{sec:phig} sumarizes the measurement of  $\BRof\BsPhiG/ \BRof\BdKstG$
and \secref{sec:bsmm} sumarizes the analysis and results of \Bqmumu.

\section{\BdToKstmm}
\label{sec:kstmm}

The rare decay \BdToKstmm is a $b \to s$, flavour changing neutral current decay, mediated by electroweak box and penguin diagrams in the Standard Model (SM). In models 
beyond the SM, new particles can enter in competing loop-order diagrams resulting in large deviations from SM predictions (see for example Refs.~\cite{Ali:1991is,Altmannshofer:2008dz}). 

\BdToKstmm candidates are selected by first applying a loose pre-selection based on the \Bd lifetime, daughter impact parameters and a requirement that the \Bd points back to one
 of the primary vertices in the event. A tighter multivariate selection, based on a boosted decision tree (BDT), is then applied to select a clean sample of \BdToKstmm candidates, with a 
signal-to-background ratio in a 100\mevcc window around the reconstructed \Bd mass of about three-to-one. The BDT is based on the \Bd kinematics, \Bd vertex quality, daughter track quality, 
impact parameter and kaon, pion and muon particle identification. The offline selection criteria are explicitly chosen to minimise angular acceptance effects. The multivariate selection 
was trained using \BdJpsiKst candidates from the 2010 data as a proxy for the signal and \BdToKstmm candidates from the upper mass sideband of the 2010 data for the background.
Specific vetoes are used in order to eliminate non combinatorial background.

The trigger, reconstruction and offline selection can all bias the measured angular distribution of \BdToKstmm candidates. The detection acceptance is accounted for by weighting 
events when fitting for $A_{FB}$, $F_{L}$ and $\mathrm{d}BF/\mathrm{d}q^{2}$ (where $q^2$ is the di-muon mass squared). Event weights are calculated on a per-event basis in a small phase space window around each candidate, 
using fully simulated Monte Carlo (MC) simulation events. 
Simulated events are re-weighted to account for known data-MC differences in PID performance, impact parameter resolution, tracking efficiency and track multiplicity.

The fit results for $A_{FB}$, $F_{L}$ and $\mathrm{d}BF/\mathrm{d}q^{2}$, and their comparison with theoretical predictions~\cite{Bobeth:2011gi}, are shown in Fig.~\ref{fig:results:lhcbonly}.

The systematic error on $A_{FB}$, $F_{L}$ and $\mathrm{d}BF/\mathrm{d}q^{2}$ is typically $\sim 30\%$ of the statistical error. In the high-$q^{2}$ 
region, the dominant contribution to the systematic uncertainty comes from the overall uncertainty on the acceptance correction  
which is dictated by the limited simulation statistics. This can 
clearly be improved for future analyses. Throughout, a sub-dominant contribution comes from the data-derived performance corrections. 
In particular, from knowledge of the PID performance and tracking 
efficiency in data.
This is again statistically limited and can also be improved with larger datasets. When fitting for $A_{FB}$ and $F_{L}$ the signal and background 
mass model and the angular model for the background 
have been varied and yield corrections at the level of 10-20\% of the statistical uncertainty. The uncertainty on the differential branching fraction 
includes the $\sim 4\%$ uncertainty coming from 
the measured \BdToJPsiKst and $\jpsi\to\mumu$ branching fractions~\cite{PDG}. These measurements are current world best, and don't
confirm previous hints of a non-SM value of $A_{FB}$ at low $q^2$.

\section{\BsPhiG}
\label{sec:phig}
In the SM, the amplitude of these $\b\to s\gamma$ penguin transitions is dominated by a virtual intermediate top quark coupling to a \W boson. Extensions of 
the SM predict new 
heavy particles that may propagate virtually within the loop  and modify the dynamics of the  transition. 
Therefore, these radiative modes are promising laboratories that could reveal the presence of new phenomena beyond the SM with the precise measurement of the 
branching ratios, asymmetries or angular distributions.
The offline selection of both the \BdKstGam and \BsPhiGam decays is performed with the strategy of  maximizing the cancellation of systematic uncertainties 
when performing the ratio. 
The analysis of $\sim 341 \invpb$ of \lhcb data gives:
\begin{equation}
\frac{\BRBdKstGam}{\BRBsPhiGam} =   1.52 \pm 0.14 \mathrm{(stat)} \pm 0.10 \mathrm{(syst)} \pm 0.12 (f_s/f_d)
\label{equation:final-result}
\end{equation}
Where $f_d$ ($f_s$) are the probabilities of the $b$ quark to hadronize into \Bd (\Bs).
This results is compatible within 1.6 standard deviations with the theory prediction.\\

Combining the ratio of branching fractions in \ref{equation:final-result} with the World Average measurement for the {\BR(\BdKstGam)} from \cite{HFAG},  we obtain,
\begin{equation}
  \BRBsPhiGam = (2.8 \pm 0.5 )\times 10^{-5}
\end{equation}
which agrees within 1.6 standard deviations with the previous experimental measuremen, and wich correspond to the most
precise measurement of this $BR$ to date.

\section{\Bqmumu}
\label{sec:bsmm}
The SM predictions for the branching fractions  
of the FCNC decays \Bsmumu 
and \Bdmumu are \BRof \Bsmumu = $(3.2 \pm0.2) \times 10^{-9}$ and \BRof \Bdmumu = $(0.10 \pm 0.01) \times 10^{-9}$~\cite{Buras2010}.
However, contributions from new processes or new heavy particles can significantly enhance these values.
For example, within Minimal Supersymmetric extensions of the SM (MSSM), 
in the large $\tan \beta$ regime,
\BRof \Bsmumu receives contributions proportional
to $\tan^6\beta$ \cite{MSSM}, where $\tan\beta$ is the ratio of the vacuum
expectation values of the two neutral \CP-even Higgs fields, and can differ significantly from the SM prediction.
The \lhcb analysis is done by clasifying \Bqmumu candidates in bins of a 2D parameter space made by the invariant mass and a multivariate clasifier which
condensates geometrical and kinematical information of the event. The signal expectation in each bin is calculated using data from control channels such
as \Bhh and \BuJpsiK. The background expectation is calculated by interpolating from mass sidebands. The \Bhh peaking background yield is calculated using
$\pi \rightarrow \mu$ and $K\rightarrow\mu$ misidentification probabilities obtained from data using decays such as \Lambdappi and $D^0\rightarrow K^+\pi^-$.
The signal and background expectations are compared with the distribution of observed events, and the limits are set using the \CLs method 
~\cite{Junk_99,Read_02}.
The \BRof \Bsmumu and \BRof \Bdmumu upper limits obtained are:
\begin{eqnarray}
\BRof{\Bsmm}  &<& 1.2 \,(1.4) \times 10^{-8}~{\rm at}~90\,\% \,(95\,\%)~{\rm CL},  \nonumber \\
\BRof{\Bdmm}  &<& 2.6 \,(3.2) \times 10^{-9}~{\rm at}~90\,\% \,(95\,\%)~{\rm CL}.  \nonumber
\end{eqnarray} 

\figref{fig:bsmm} shows the luminosity needed to impose stronger limits or to achieve a 3$\sigma$ evidence of \Bsmumu. 

\section{Conclusions}
\label{conc}
\begin{figure}[H]
\centering
\includegraphics[width=0.4\textwidth]{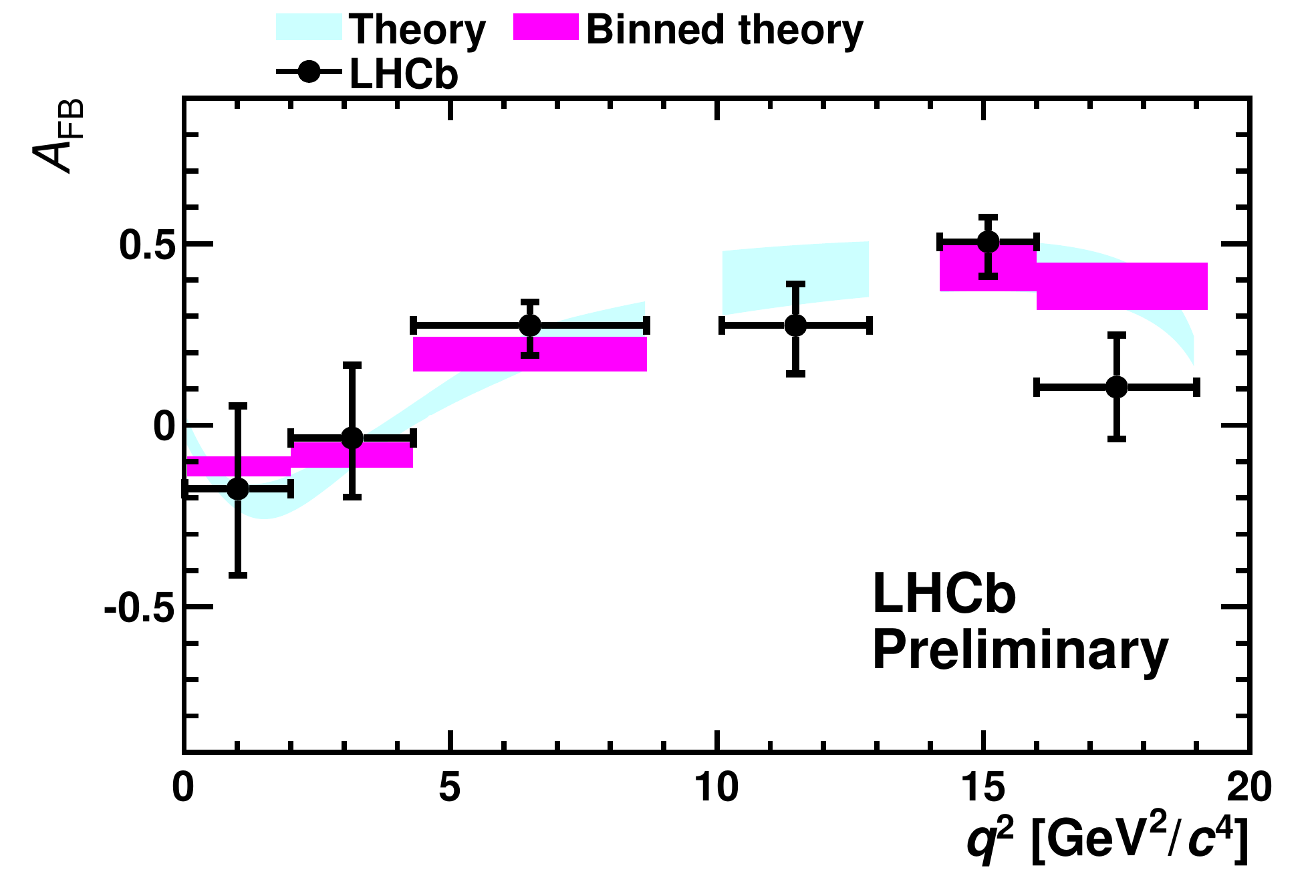}
\includegraphics[width=0.4\textwidth]{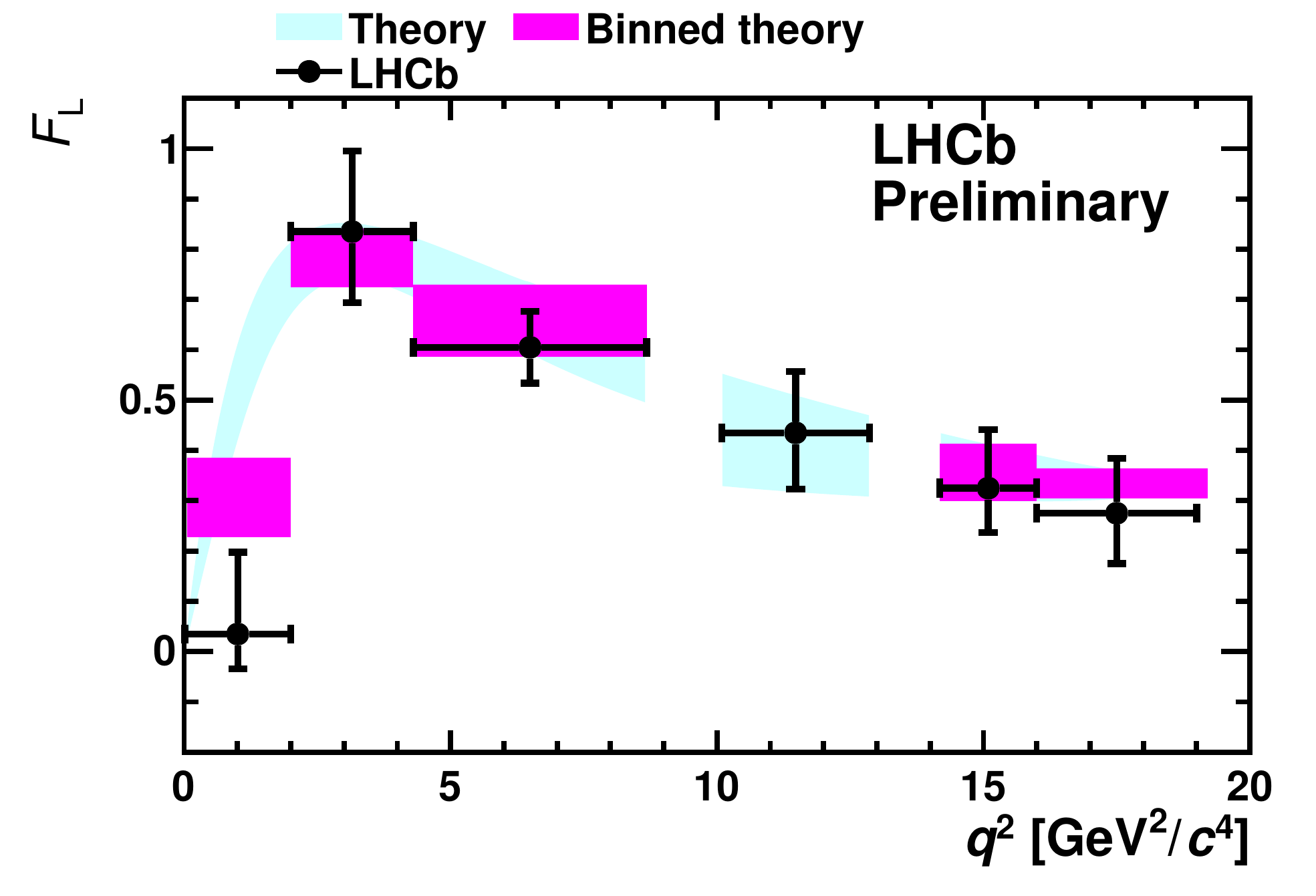}
\includegraphics[width=0.4\textwidth]{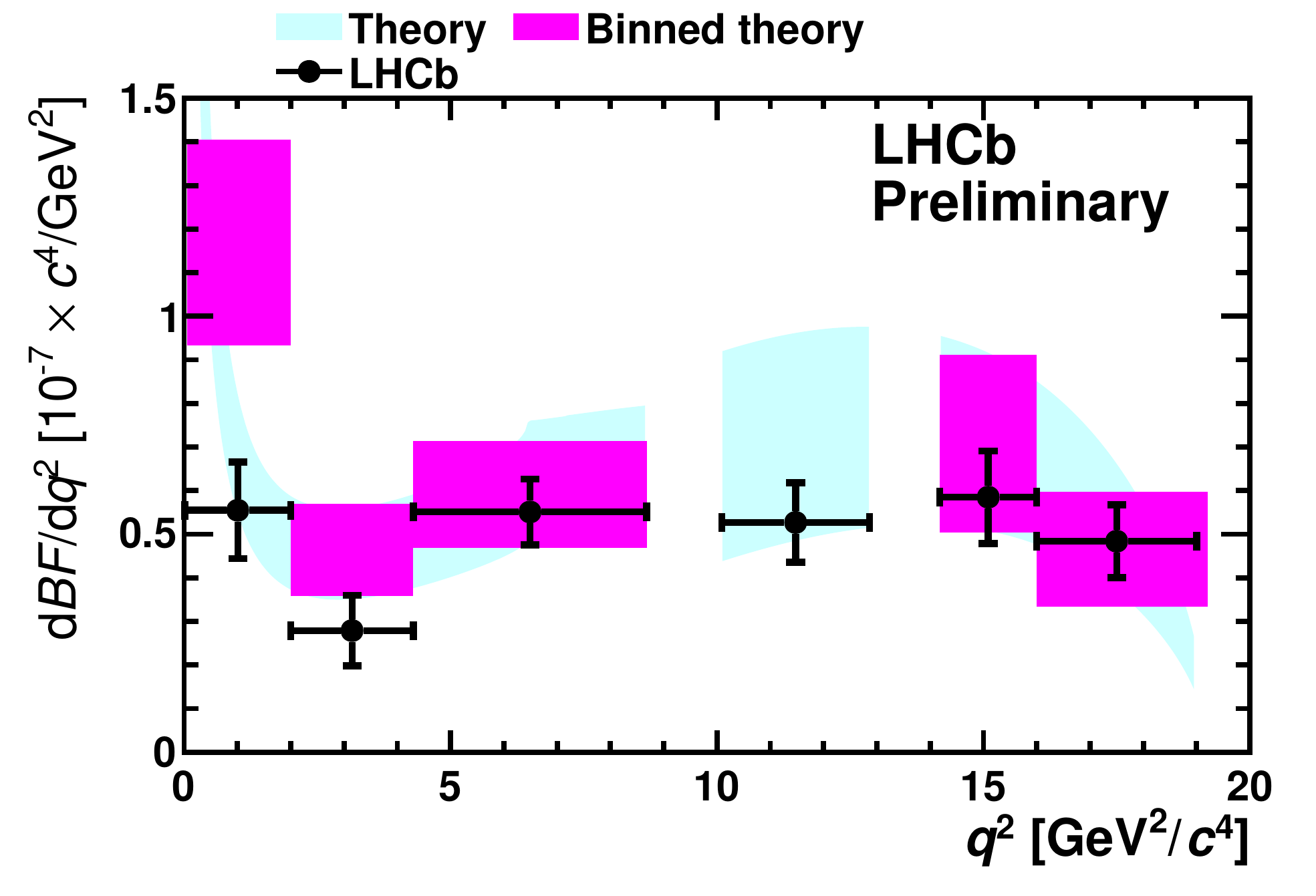}
\caption{
$A_{FB}$, $F_{L}$ and the differential branching fraction as a function of $q^{2}$ in the six Belle $q^{2}$ bins. The theory predictions are described from Ref.~\cite{Bobeth:2011gi}.
\label{fig:results:lhcbonly}
}
\end{figure}

As can be seen in Fig.~\ref{fig:results:lhcbonly}, there is good agreement between recent SM predictions and \lhcb's measurement of $A_{FB}$, $F_{L}$ and $\mathrm{d}BF/\mathrm{d}q^{2}$ in the 
six $q^{2}$ bins. In a $1 < q^{2} < 6 \gev^{2}$ bin, \lhcb measures $A_{FB} = -0.10^{+0.14}_{-0.14}\pm 0.05$, $F_{L} = 0.57^{+0.11}_{-0.10} \pm 0.03$ and 
$\mathrm{d}BF/\mathrm{d}q^{2} = 0.39\pm0.06\pm0.02$, to be compared with theoretical predictions of $A_{FB} = -0.04^{+0.03}_{-0.03}$, $F_{L} = 0.74^{+0.06}_{-0.07}$ and 
$\mathrm{d}BF/\mathrm{d}q^{2} = (0.50^{+0.11}_{-0.10}) \times 10^{-7}$ respectively. The experimental uncertainties are presently statistically dominated, and will 
improve with a larger data set. Such a data set would also enable \lhcb to explore a wide range of new observables ~\cite{Egede:2008uy}. 

In 340 pb$^{-1}$ of $pp$ collisions at a centre of mass energy of $\sqrt{s}=7$~TeV the most precise measurement of \BRof\BsPhiGam has been performed, giving:
\begin{equation}
\frac{\BRBdKstGam}{\BRBsPhiGam} =   1.52 \pm 0.14 \mathrm{(stat)} \pm 0.10 \mathrm{(syst)} \pm 0.12 (f_s/f_d)
\end{equation}

The \BRof \Bsmumu and \BRof \Bdmumu upper limits obtained by LHCb are:
\begin{eqnarray}
\BRof{\Bsmm}  &<& 1.2 \,(1.4) \times 10^{-8}~{\rm at}~90\,\% \,(95\,\%)~{\rm CL},  \nonumber \\
\BRof{\Bdmm}  &<& 2.6 \,(3.2) \times 10^{-9}~{\rm at}~90\,\% \,(95\,\%)~{\rm CL}.  \nonumber
\end{eqnarray} 

\begin{figure}[H]
\centering
\includegraphics[width=0.4\textwidth]{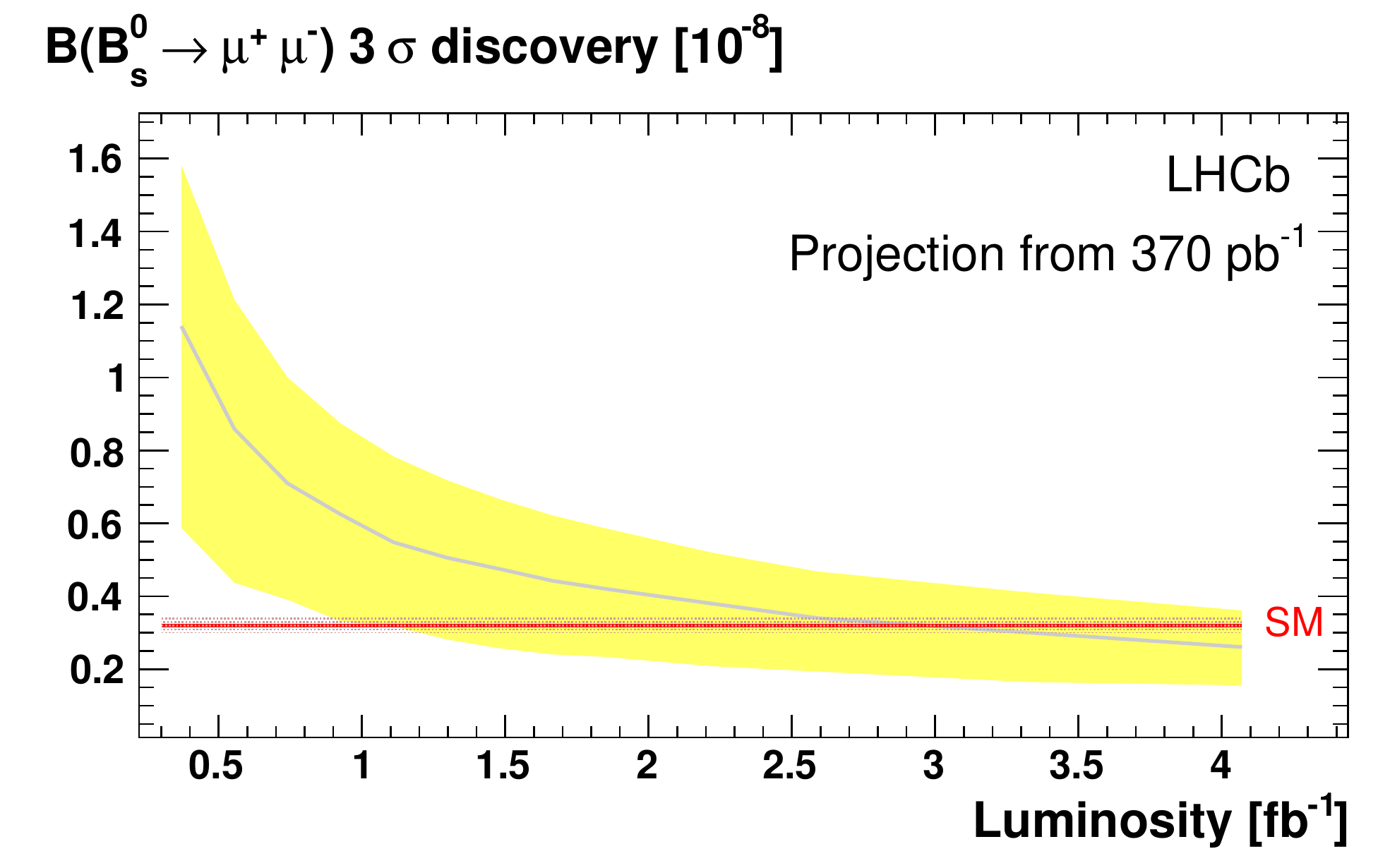}
\includegraphics[width=0.4\textwidth]{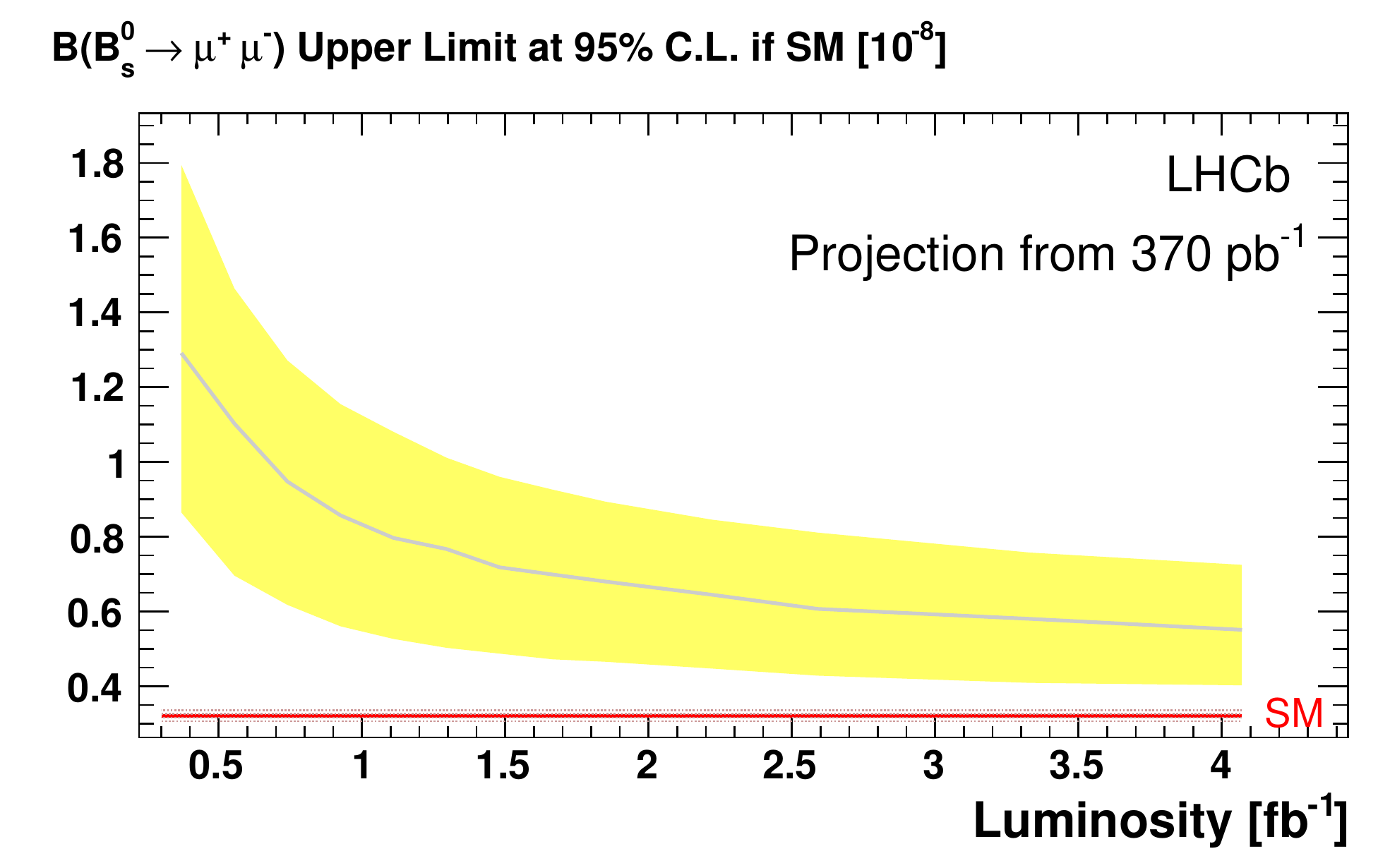}
\includegraphics[width=0.4\textwidth]{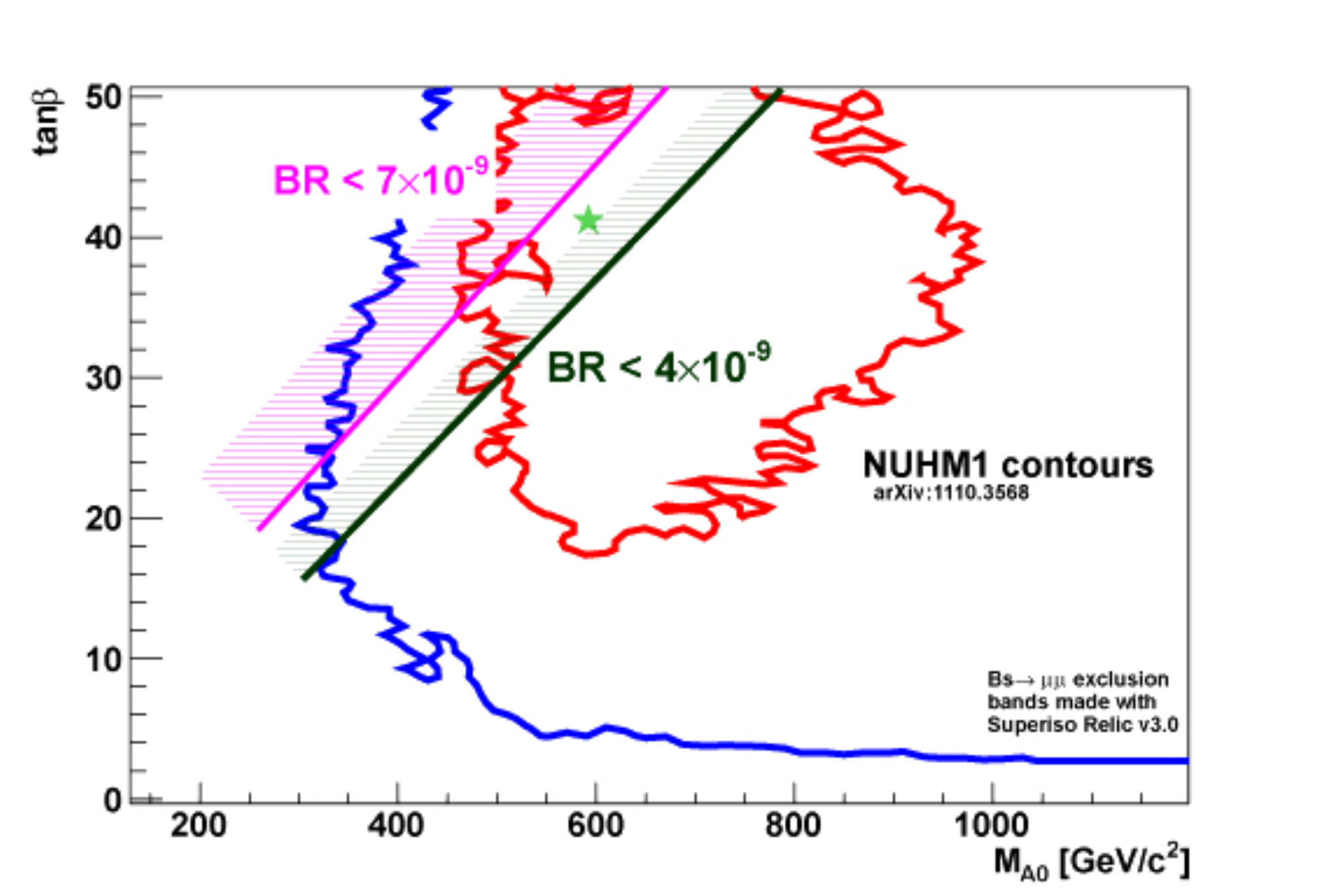}
\caption{
Luminosity needed in order to get a \Bsmm 3$\sigma$ evidence (top) or a $95\%CL$ exclusion in the presence of a SM signal (center).
The bottom plot shows how upper limits in the $10^{-9}$ level would constraint the region around the minimum of the NUHM1 fit
from ~\cite{MC7}.
\label{fig:bsmm}
}
\end{figure}

In \figref{fig:bsmm} the luminosity needed for a 3$\sigma$ evidence as a function of \BRof\Bsmumu is shown.
Approximately $\sim 2\invfb$ are needed in the case that the value is equal to the SM prediction, but statistical
fluctuations can make it possible with $\sim 1\invfb$.
\figref{fig:bsmm} also shows that exclusions of \BRof\Bsmumu down to the ($2\times)$ SM level would
impose important constraints in region around the current NUHM1 best fit point ~\cite{MC7}.
All the results presented here are current world best.

\end{document}